\documentclass[final,authoryear, 3p,times,twocolumn]{elsarticle}
\usepackage[colorlinks=true,linkcolor=black, citecolor=blue, urlcolor=blue]{hyperref}

\usepackage{amssymb}
\usepackage{algorithm}
\usepackage{algpseudocode}
\usepackage{graphicx}	
\usepackage{amsmath}	
\usepackage{subcaption}
\usepackage{tikz}
\usepackage{scalerel}
\usepackage{comment}
\usepackage{orcidlink}
\usepackage{colortbl}
\usepackage{array,multirow}
\usepackage{xspace}

\journal{Astronomy and Computing}

\newcommand{\scarlet}{{\sc scarlet}\xspace}
\newcommand{\scarlettwo}{{\sc scarlet\oldstylenums{2}}\xspace}
\newcommand{\bx}{\mathbf{x}}

\begin{document}
\begin{frontmatter}

\title{Score-matching neural networks for improved multi-band source separation}

\author[inst1]{Matt L. Sampson$^{\orcidlink{0000-0001-5748-5393}}$}
\ead{matt.sampson@princeton.edu}
\author[inst1,inst2]{Peter Melchior$^{\orcidlink{0000-0002-8873-5065}}$}
\author[inst1]{Charlotte Ward$^{\orcidlink{0000-0002-4557-6682}}$}
\author[inst1]{Sufia Birmingham$^{\orcidlink{0009-0003-7044-9751}}$}

\affiliation[inst1]{organization={Department of Astrophysical Sciences, Princeton University},
            city={Princeton},
            postcode={08544}, 
            state={NJ},
            country={USA}}

\affiliation[inst2]{organization={Center for Statistics and Machine Learning, Princeton University},
            city={Princeton},
            postcode={08544}, 
            state={NJ},
            country={USA}}

\begin{abstract}
We present the implementation of a score-matching neural network that represents a data-driven prior for non-parametric galaxy morphologies.
The gradients of this prior can be incorporated in the optimization of galaxy models to aid with tasks like deconvolution, inpainting or source separation. We demonstrate this approach with modification of the multi-band modeling framework \scarlet that is currently employed as deblending method in the pipelines of the HyperSuprimeCam survey and the Rubin Observatory. The addition of the prior avoids the requirement of non-differentiable constraints, which can lead to convergence failures we discovered in \scarlet. We present the architecture and training details of our score-matching neural network and show with simulated Rubin-like observations that using a data-driven prior outperforms the baseline \scarlet method in accuracy of total flux and morphology estimates, while maintaining excellent performance for colors. We also demonstrate significant improvements in the robustness to inaccurate initializations. The trained score models used for this analysis are publicly available at \url{https://github.com/SampsonML/galaxygrad}.
\end{abstract}

\begin{keyword}
methods: machine learning, data analysis \sep techniques: image processing
\end{keyword}

\end{frontmatter}



\section{Introduction} \label{sec:intro}
Upcoming astronomical surveys will produce huge amounts of data, with current estimates for the Rubin Observatory Legacy Survey of Space and Time \citep[LSST;][]{LSST_2019} yielding on the order of 10 billion galaxies. This vast amount of data is meant to greatly reduce the statistical uncertainty of cosmological and astrophysical parameters. LSST specifically covers about half the sky to a limiting magnitude of $i \approx 27$. While detecting and measuring more faint and distant galaxies than ever before, this sensitivity increase will also drastically increase the occurrence of overlap between neighboring sources, a systematic effect known as \textit{blending}. Galaxy blending has already been shown to affect up to 60\% of galaxies in the Hyper-Suprime Cam (HSC) wide survey \citep{bosch2018} that has a limiting magnitude of $i \approx 26$. Blending thus constitutes a leading systematic issue in gravitational lensing and galaxy clustering studies \citep{mandelbaum_2018,melchior_blend_2021,Nourbakhsh_2022}. Therefore, accurate and computationally efficient deblending of galaxies, or an effective calibration scheme that captures the effects of blending \citep{, MacCrann_2022}, is crucial to allow us to maximize the value of upcoming astronomical surveys. We pursue the former approach so as to minimize the efforts that may still be required for the latter.

There are a variety of publicly available deblending codes, including \textsc{SExtractor} \citep{sextractor_1996}, the SDSS deblender \citep{Lupton2005SDSSIP} and \textsc{muscadet} \citep{muscadet_2016}, and \scarlet \citep{scarlet_2018}. 
Methods based on neural network, usually trained on simulated or reliably unblended galaxies, have recently also shown good performance \citep[e.g.][]{gan_deblend_2019, burke_2019_RCNN, Boucaud2020-xt, Arcelin_VAE_2021, gan_hemmati_2022, Wang_2022_deblend_RNN}.
In this work, we combine the strengths of the \scarlet method, which is currently implemented in the pipelines for HSC and LSST, with the flexibility of neural networks.

We describe our motivation and the neural score model in \autoref{sec:methods}. In \autoref{sec:model_perform} we demonstrate the capability of the score model to represent galaxy morphologies. We compare the performance of the original \scarlet implementation to the neural-network enhanced version, on real and simulated data in \autoref{sec:scarlet2_metrics}, and summarize our findings in \autoref{sec:summary}.

\section{Methodology} 
\label{sec:methods}

\subsection{Source separation}
\label{sec:source}
\begin{figure*}[t]
    \includegraphics[width=0.98\textwidth]{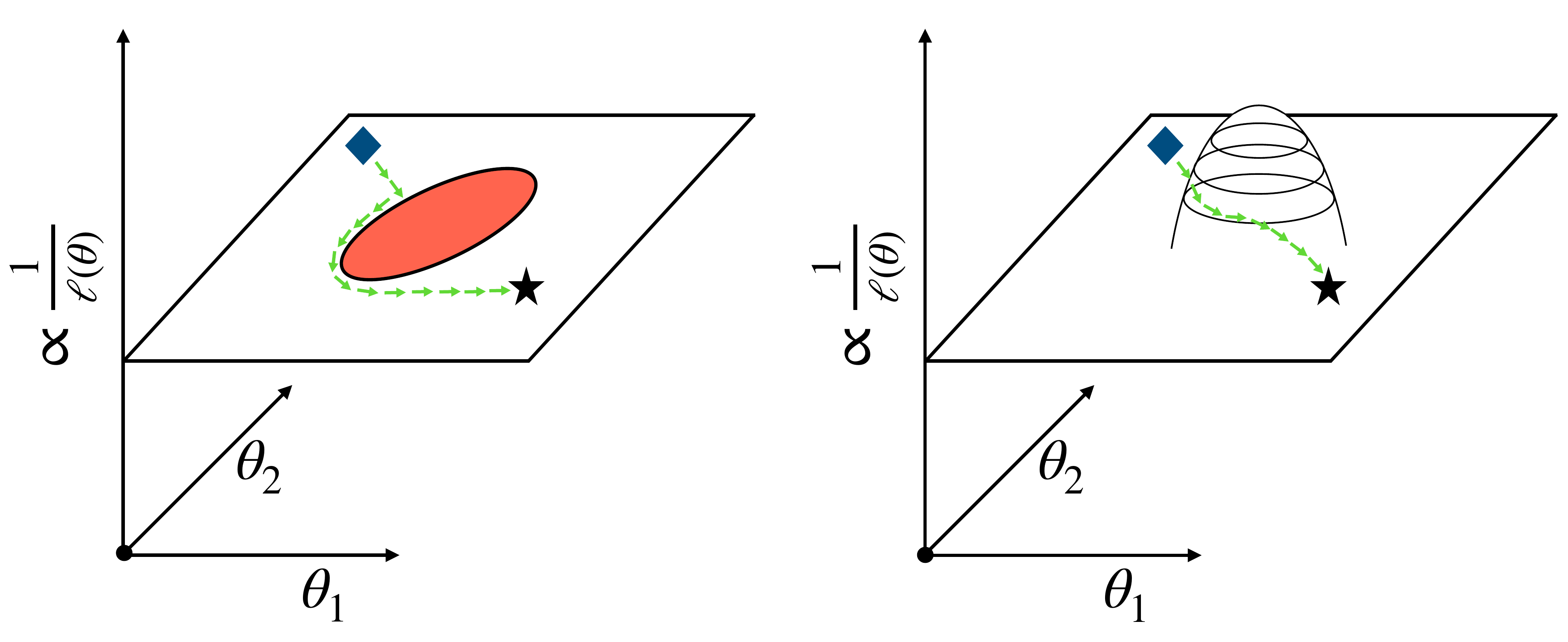}
    \caption{Sketch of optimization paths. \scarlet (\emph{left}) uses constrained optimization to enforce physically desirable heuristics for galaxies, such as non-negativity and monotonic radial profiles, by proximal projections, which prevent the optimizer from taking steps through infeasible regions, slowing down or even impeding convergence. \scarlettwo (\emph{right}) uses data-driven priors, which provide a continuous and fully differentiable posterior surface.}
    \label{fig:constraints}
\end{figure*}

\scarlet separates and reconstructs astronomical sources in \textit{multi-band} images through a generalized method of non-negative matrix factorization termed Constrained Matrix Factorization, where for every source $k$ one factor, $\mathbf{A}_k$, refers to the spectral energy distribution across bands and the other, $\mathbf{S}_k$, to the intensity variation across the sky. An astronomical scene with a hyperspectral cube $\mathbf{Y}$ is model as a linear combination of multiple sources,
\begin{equation}
\label{eq:sum-sources}
\mathbf{Y} = \sum_k \mathbf{A}_k \times \mathbf{S}_k.
\end{equation}
 
 In this method, physically motivated parameter constraints are added to the optimization routine by means of proximal operators \citep{Combettes2008-lv, Combettes2009-ew, Parikh2014-ex} to ensure that the individual deblended sources appear realistic \citep[see sec 2.3 of][for details]{scarlet_2018, Moolekamp2018-kd, Melchior2019-be}. Doing so is necessary for source separation or other ill-defined inverse problems because the likelihood function alone cannot distinguish between the many degenerate solutions that arise with overlapping galaxies, most if which will not look like galaxies. In \scarlet source  constraints  ensure that the reconstructed pixel value never falls below zero and that the radial profile decreases monotonically. This method is fast and provides a robust and effective non-parametric galaxy model, but many galaxies may have locally non-monotonic behavior, e.g. for tightly wound spiral arms. A more flexible model would be beneficial to better capture such features.

The inflexibility of those constraints occasionally even leads to unrealistic features, such as radial streaks, that formally satisfy the constraints.
Such issues originate in the proximal gradient method, which finds the minimizer of $f(\bx)+g(\bx)$, where $f$ is a convex, differentiable function, in our case the negative log-likelihood function, and $g$ is a regularizer, by the sequence
\begin{equation}
\bx^{t+1} \leftarrow \text{prox}_{\lambda^t g}\left(x^t- \lambda^t\nabla f(x^t)\right)
\end{equation}
with step sizes $\lambda^t$.
For a non-differentiable regularizer, its proximal operator $\text{prox}_{\lambda g}$ amounts to a projection onto the surface of the manifold that satisfies the constraint, a so-called ``subgradient''.
If the gradient vector is directly opposed to the subgradient, this sequence leads to very slow convergence or even to a trapping of the optimizer in locations with very large $f$ (see \autoref{fig:constraints} for a sketch).

These problems arise in practice. We found \scarlet to be vulnerable to getting trapped if, due to a bad initialization, the galaxy model requires large changes in the overall flux of a source, which leads to gradients with pronounced oscillating ring-like patters and thus reconstructions with negative pixel values. But these are not allowed for non-negative models and therefore rejected by the proximal operator $\mathrm{prox}_+(\bx) = \max(\bx, 0)$. The most direct gradient path is thus blocked (see \autoref{fig:bad_init} for an example and \autoref{sec:init} for a discussion of this failure mode). It is important to realize that the proximal projection is applied after every gradient step, so that this trapping can occur \emph{during} the optimization even if the final model would satisfy the constraints because an intermediate model might violate them.

Lastly, non-differentiable regularizers also preclude most continuous methods, e.g. to determine parameter uncertainties through second derivatives of the log-likelihood function, or sampling methods. We thus seek to replace proximal optimization of the constrained likelihood,
\begin{equation}
\label{eqn:optimise}
    L(\bx) =  \log\mathcal{L}(\bx \mid \mathcal{D}) + g(\bx),
\end{equation}
with regular optimization of the Bayesian posterior,
\begin{equation}
\label{eqn:optimise2}
    L(\bx) =  \log\mathcal{L}(\bx \mid \mathcal{D}) + \log \mathcal{P}(\bx),
\end{equation}
wherein the prior distribution $\mathcal{P}(\bx)$ takes the place of the constraints. In this work we focus on priors on galaxy morphologies $\mathbf{S}_k$.

\subsection{Model for the prior distribution}
\label{sec:prior}

Score-matching diffusion models are a class of generative deep learning models that produce high-quality sample images from a prior distribution. This is achieved through a neural network that is trained to approximate the gradient of the logarithmic prior distribution $\nabla \log \mathcal{P}(\mathbf{x})$, also known as the \textit{score function}. There has been a great amount of work recently with these types of models, primarily in the domain of generative image modeling \citep{song2019generative,song_2020_improve,Ho_DDPM_2020,song_time_2020,Nichol_DDPM_2021} as well as adaptations to astronomy to generate realistically looking galaxies \citep{score_galaxies_2021}. 

We utilize the same architecture presented in \citet{song_time_2020} and use the same loss function with minimal alterations. In this implementation of a diffusion model, the intended goal is to attain a solution to the It$\hat{o}$ stochastic differential equation (SDE), called the \textit{forward process},
\begin{equation}
\label{eqn:fwd_diffusion}
    \mathrm{d}\bx = \mathbf{f}(\bx,t)\mathrm{d}t + g(t)\mathrm{d}\mathbf{w},
\end{equation}
where $\bx\sim\mathcal{P}(\bx)$ is a sample from the data-generating process, $\mathbf{f}$ is a deterministic drift function, and $\mathbf{w}$ is a sample from a standard Wiener process. For any time $t\in[0, T]$, the forward diffusion process amounts to a mean drift and an increasing addition of noise, both of which can easily be simulated. \citet{anderson1982reverse} proved the existence of an analytical form to reverse \autoref{eqn:fwd_diffusion} as
\begin{equation}
\label{eqn:bwd_diffusion}
    \mathrm{d}\bx = \left[ \mathbf{f}(\bx,t) - g^2(t) \nabla \log \mathcal{P}_t(\bx) \right]\mathrm{d}t + g(t) \mathrm{d}\overline{\mathbf{w}},
\end{equation}
where $\mathcal{P}_t$ is the \emph{time-dependent} form of the prior distribution and $\overline{\mathbf{w}}$ is again a standard Wiener process sample. 
In other words, when starting with noisy samples $\bx(T)\sim\mathcal{P}_T(\bx)$ and reversing the diffusion, we can sample from $\bx(0)\sim\mathcal{P}(\bx)$, our desired target distribution.
Intuitively, any well-guided attempt of removing noise from the result of a diffusion process requires knowledge of the distribution of likely samples.

\autoref{eqn:bwd_diffusion} requires the evaluation of $\nabla\log\mathcal{P}_t(\mathbf{X})$, which we seek to accelerate by training a neural network $s_\theta$ on the following loss:
\begin{equation}
\label{eqn:loss}
\begin{split}
    L(\theta) = &\mathbb{E}_{\mathbf{x}(0)}\mathbb{E}_{\mathbf{x}(t)\mid\mathbf{x}(0)} \Big( \big\lVert s_{\theta}(\mathbf{x}(t),t)\\
    &- \nabla_{\mathbf{x}(t)} \log \mathcal{P}_{0t}(\mathbf{x}(t) \mid \mathbf{x}(0))\big\rVert^2_2 \Big).
\end{split}
\end{equation}
The specific form of the time-dependent prior depends on $\mathbf{f}$ and $g$. 
\citet{song_time_2020} demonstrated excellent performance for a sequence of noise scales $\beta(t)$ with
\begin{equation}
\begin{split}
&\mathbf{f}(\bx, t) = -\frac{1}{2}\beta(t)\bx \ \mathrm{and}\\
&g(t) = \sqrt{\beta(t)(1-\mathrm{e}^{-2\int_0^T\beta(s)\mathrm{d}s})},
\end{split}
\end{equation}
which leads to the time-dependent prior
\begin{equation}
\begin{split}
\mathcal{P}_{0t}(\mathbf{x}(t) \mid \mathbf{x}(0)) &= \mathcal{N}\left(\mathbf{x}(t) \mid \mathbf{m}(t), \mathbf{V}(t)\right)\ \mathrm{with}\\
\mathbf{m}(t) &= \bx(0)\,\mathrm{e}^{-\frac{1}{2}\int_0^T\beta(s)\mathrm{d}s}\ \mathrm{and}\\
\mathbf{V}(t) &= 1-\mathrm{e}^{-\int_0^T\beta(s)\mathrm{d}s}\mathbf{I}.
\end{split}
\end{equation}
Once the integral $\int_0^T\beta(s)\mathrm{d}s$ is evaluated, the gradients of logarithm of this prior can be analytically calculated and inserted in \autoref{eqn:loss} to train the score-matching diffusion model $s_\theta$. 
We note here that while having a fully generative model might be useful in other contexts, we only use an estimate of the gradient of the log-prior for our optimization routine. Hence we choose a score-matching diffusion model to directly estimate this quantity instead of working with alternative generative models.

\subsection{Full implementation}

To test the utility of the score model to aid source separation, we implement the methods described in \autoref{sec:source} and \autoref{sec:prior} as a reformulation of the \scarlet method of \citet{scarlet_2018}. The new code, dubbed \scarlettwo\footnote{\url{https://github.com/pmelchior/scarlet2}}, is implemented in \texttt{jax} and \texttt{equinox} and hence compatible with CPU, GPU and TPU devices with no code alteration required. While the purpose of the new code is to enable large-scale production runs for futures surveys as well as custom investigations of smaller data sets, for this work we only make use of it to demonstrate and denote prior-based compared to the existing constraint-based optimization for the same model specification of \autoref{eq:sum-sources}.

\section{Diffusion model training}
\label{sec:model_perform}

\begin{figure*}[t]
    \includegraphics[width=0.98\textwidth]{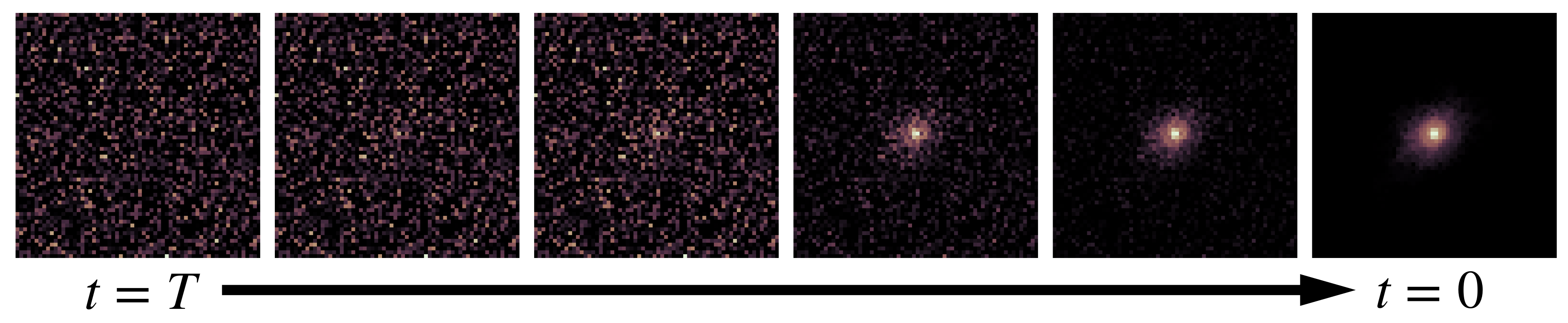}
    \caption{Sample from the HSC galaxy morphology prior. The left most panel shows the initial random Gaussian noise field, from which noise is removed step by step. The final denoised image at time $t = 0$ is then a sample consistent with the data distribution.}
    \label{fig:diffusion_sample}
\end{figure*}

We train our neural network model to reverse the diffusion process from the arbitrary start point of $T = 10$ until to the zero-temperature (sometimes \textit{zero-time}) end point of $t = 0$. We use a noise scheduling of $\sigma = \sqrt{(1 - e^{-T})}$ for $T \in \{10, 0\}$, where $\sigma$ acts to scale the pixel values of a 2D array of pure Gaussian random noise. A demonstration of the full reverse diffusion process is shown in \autoref{fig:diffusion_sample}, and the details of the training process are described below.

\subsection{Datasets}
\label{sec: dataset}
For our initial training set of galaxies we used the recent source catalog from HSC. \scarlet has already been run on the entirety of this catalogue, we therefore have direct access to the deblended sources. We extracted around 250,000 sources from each patch for three tracts $\{9615, 9697, 9813\}$. The sources are centered in their respective images, with image dimensions ranging from $15\times 15$ to $221\times 221$ pixels. 
For our diffusion model, we decided to train on only 2 sizes, \textit{low resolution} of size $32 \times 32$ and \textit{high resolution} $64 \times 64$. For images larger than $64 \times 64$, there are far fewer observations, and at these resolutions galaxies are usually highly significant and therefore rarely need to regularization from a prior. To construct our two training sets we use zero-padding around the edges of the original source images.

We are aware that defining a prior based on the \scarlet models carries the risk of re-introducing its problems into \scarlettwo. However, prominent failures in \scarlet are not common, and posterior optimization is more permissive than constrained optimization. We therefore expect that inaccuracies from the prior can be overcome by data. In principle, one could retrain the prior with the outputs from our new model and bootstrap a highly accurate informative prior this way \citep{Rozet2024-gm,Barco2024-bz}.

We also train a prior model on a mock images mimicking Zwicky Transient Facility (ZTF) observations \citep{Bellm2019,Graham2019,Dekany2020TheSystem}. For this, we create a range of isolated galaxies using \texttt{Galsim} \citep{2015galsim} Sersic profiles of $n=1$ (exponential profile) and $n=4$ (De Vaucouleurs profile). We vary four morphological parameters of the simulated galaxies: the radius, axis ratio, rotation angle, and sub-pixel shift from image center, with probabilities determined by the distribution of host galaxy properties from the sample of 1626 variable AGN in ZTF presented in \citet{Ward2023}. We simulate 10,000 galaxies (5,000 galaxies for both Sersic flux profiles) each in two box sizes of $32^2$ and $64^2$ pixels.

\subsection{Data transformation}
The majority of our training data for both the HSC and the mock ZTF images often comprise of only a small amount of bright pixels in the central region of the image, with the rest of the image being close to zero. 
To enhance the low brightness pixels, we applied a transformation
\begin{equation}
    h(\mathbf{x}) = \frac{\log(\mathbf{x} + 1)}{0.1},
\end{equation}
which we found to result in increased model performance, especially for faint galaxy features.
To undo this transform, and ensure that the gradient used in \scarlettwo's optimisation routine is in the same coordinates of the observation, we add the Jacobian of the transformation to the score function,  
\begin{equation}
 \nabla \log P_\mathbf{x} \to \nabla \log P_\mathbf{x} \cdot \mathbf{\mathrm{J}}(h),
\end{equation}
where the calculation of the Jacobian of $\mathbf{\mathrm{J}}(h)$ is done by automatic differentiation in \texttt{jax}.

\subsection{Generative capabilities}
\begin{figure*}[t]
    \centering
    \includegraphics[width=0.97\textwidth]{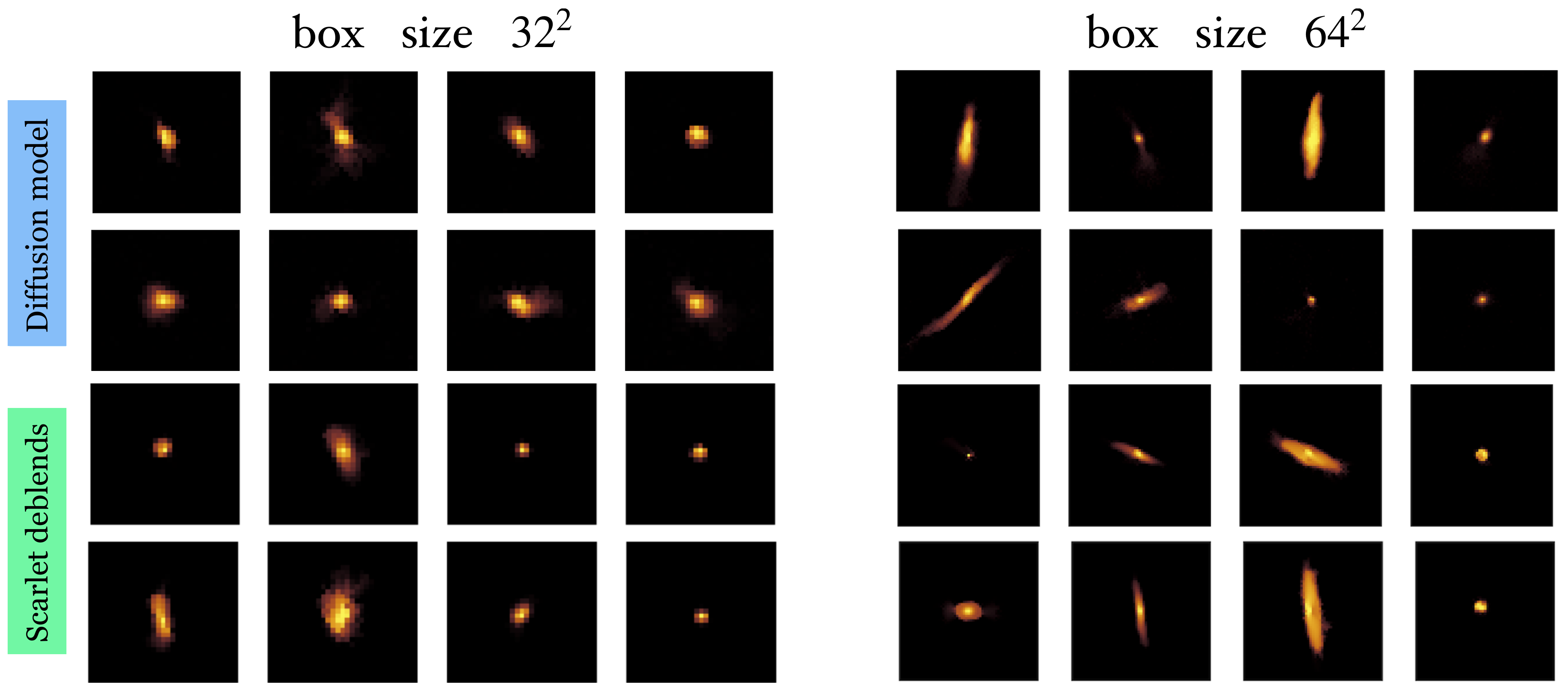}
    \caption{Comparison between generated samples, and randomly chosen samples from the training set for both size $32^2$ and $64^2$ images.}
    \label{fig:samples}
\end{figure*}

An example of the trained diffusion process is shown in \autoref{fig:diffusion_sample}, which is based on the HSC prior. The denoising process takes place over an arbitrary time period $t=T \to t=0$, with a noticeable galaxy center appearing approximately halfway into the process. Additional diffusion-model generated examples for both image sizes are shown in \autoref{fig:samples}. Visually, there is no discrepancy between the generated samples and the training data, consistent with other studies showing the successful performance of diffusion models to generate realistic galaxy images \citep{score_galaxies_2021, Adam2022}. 

\subsection{Out-of-domain performance}
\begin{figure}[t]
    \includegraphics[width=0.47\textwidth]{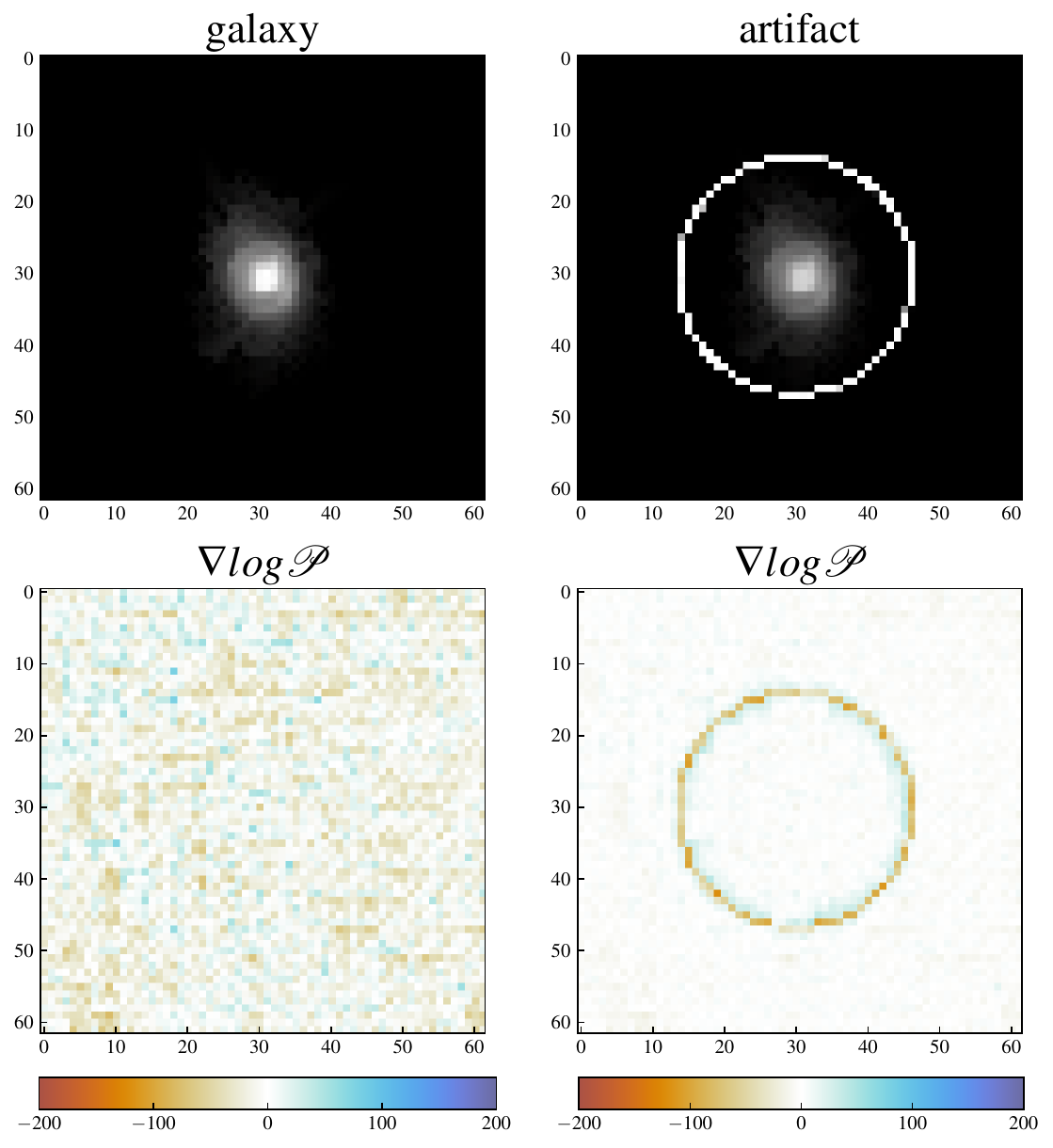}
    \caption{Artifact removal. A galaxy sampled from our morphology prior (\emph{top left}) shows flat prior gradients (\emph{bottom left}), but an added ring artifact (\emph{top right}) causes large gradients that seek to suppress the features that do not occur in the training data (\emph{bottom right}).}
    \label{fig:artifacts}
\end{figure}

Our model is trained to generate the gradients of a log-prior distribution of individual galaxies. One would therefore expect that, when shown images of objects that clearly deviate from the galaxy samples, the gradient field would seek to suppress features that are unlike those seen in galaxies. To test this, we take the generated sample from \autoref{fig:diffusion_sample} and add an artifact in the form of a ring around the galaxy, as shown in the top row of \autoref{fig:artifacts}. We then evaluate the score function on both the original and altered image, which is shown in the bottom row of \autoref{fig:artifacts}. The gradients are evaluated at the zero-temperature point ($t = 0$) because that corresponds to the gradient field of the final denoised sample. We see in the gradients of the unaltered image that there is no clear structure, which is expected as this image represents a direct sample from the prior. 

In the image with the artifact, we clearly see that the ring feature is strongly suppressed by the prior with high negative values in the gradients at the precise location of the ring.

This tests demonstrates not only that the data-driven galaxy prior is useful to render the models robust to artifacts but also that the main features of the gradient are stable to tests outside of the training domain.
The ability to suppress out-of-domain features should be particularly useful for source separation tasks, where the light contamination of a nearby source will appear atypical given the \emph{isolated} galaxies in our training data.

 \section{Prior-informed source separation} 
 \label{sec:scarlet2_metrics}

To deblend sources, \scarlet employs a first-order gradient optimization scheme, which computes the gradients of the loss function \autoref{eqn:optimise} with an explicit likelihood, e.g. for uncorrelated Gaussian noise,
\begin{equation}
\log\mathcal{L} =  -\frac{1}{2}\left\lVert \mathbf{W}\cdot\left(\mathbf{Y} - \mathbf{P}\sum_k \mathbf{A}_k\times\mathbf{S}_k\right)\right\rVert_2^2,
\end{equation}
where $\mathbf{Y}$ is a multi-band image cube, $\mathbf{W}$ the inverse variance weights in every pixel, and $\mathbf{P}$ is a, typically linear, forward operator that describes the instrument response.
Following the approach of \citet{Lanusse2019}, \scarlettwo optimizes \autoref{eqn:optimise2}, which maintains the likelihood gradient and thus a full forward model of the instrument, and adds priors for any source morphology $\mathbf{S}_k$:
\begin{equation}
\nabla_{\mathbf{S}_k}L = \nabla_{\mathbf{S}_k}\log\mathcal{L}(\mathbf{A}_1,\dots,\mathbf{A}_K, \mathbf{S}_1, \cdots, \mathbf{S}_K\mid\mathbf{Y}) + s_\theta(\mathbf{S}_k)
\end{equation}
While one could add additional priors to the source SEDs, morphologies exhibit usually more degeneracies in the region of source overlap, so that morphology priors are more critically needed.

To be clear, we only evaluate the score function $s_\theta$ for each $\mathbf{S}_k$, not the prior value itself, as doing the latter would be very costly. We therefore do not obtain an estimate of the log-posterior value, but we can be assured that by following these gradients (either through gradient descent or in a Hamiltonian Monte Carlo approach) we find the maximum of (or sample from) the posterior.

\subsection{Zero-temperature evaluation}
Running a full diffusion model to sample from the data distribution is computationally costly as it often requires hundreds of calls to the score function. However, for our purposes, we only wish to call an evaluation of the score function at $t = 0$, i.e. the gradient of the original data distribution. If possible, doing so would save significant computation costs, but the entire development of diffusion model was motivated by the insight that such zero-temperature gradients are usually inaccurate \citep{song2019generative}. So, why do we believe that we can get away with it? Because 1) galaxy morphologies are not nearly as complex as natural images which form the usual test cases and have shown to noticeably benefit from the diffusion process, and 2) we initialize each source with a reasonable first guess to the true morphology shape, so that we usually stay close to the range of examples the score-model has been trained on. This is opposed to initializing with an array of Gaussian noise, as is the standard for a full diffusion model. \autoref{fig:artifacts} even demonstrates that out-of-distribution artifacts have valid zero-temperature gradients, which implies that argument 1) above is more important than argument 2), which inspires our initial confidence that a full diffusion process would be mostly unnecessary for source separation of the majority of galaxies in large survey. We will now test this approach and note that a full diffusion model, where one would take $n$ steps from $T=10 \to 0$ for each likelihood update, might be required for more complicated morphologies. 

\subsection{Example}
\begin{figure*}[h!]
    \centering
    \includegraphics[width=\textwidth]{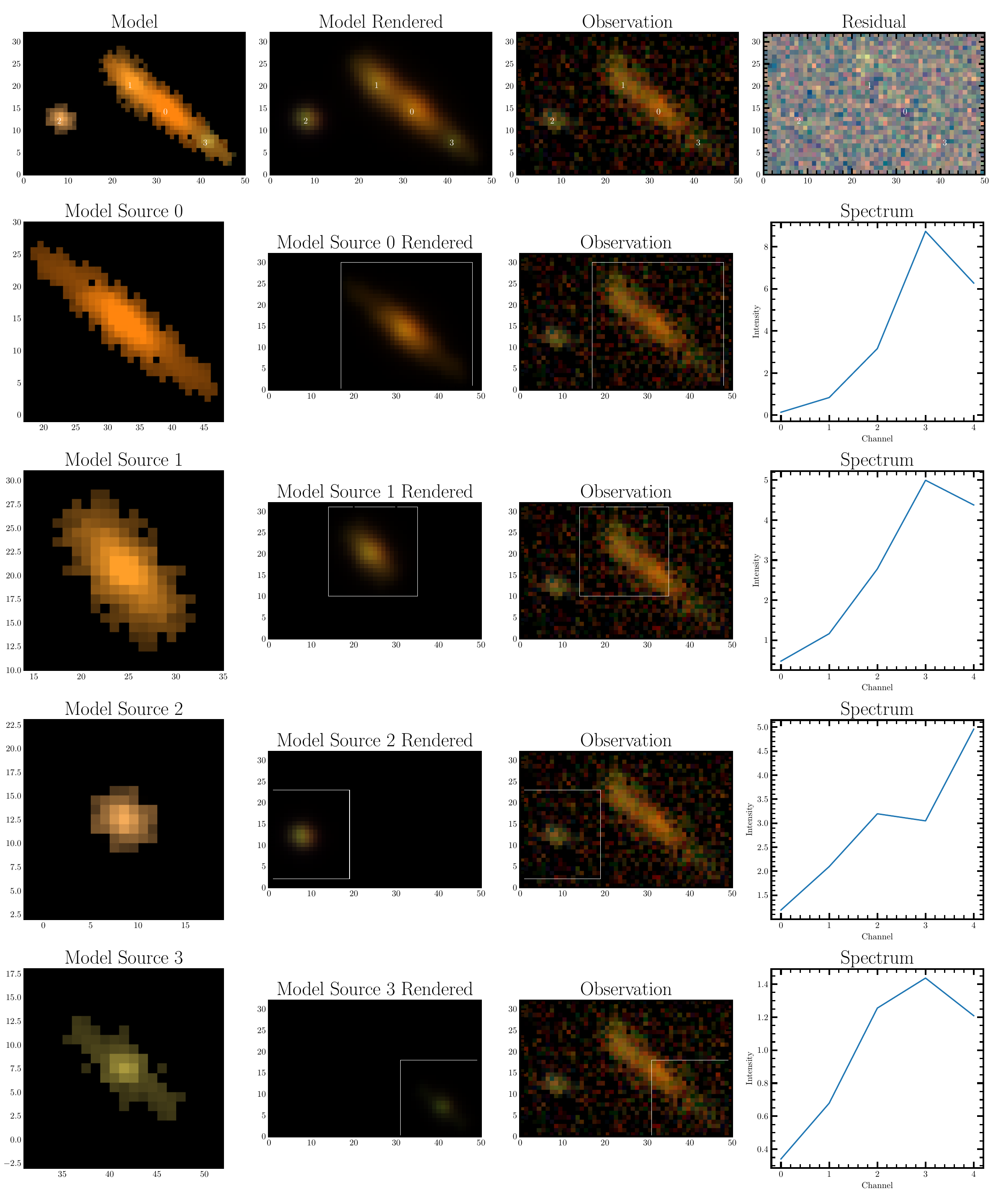}
    \caption{Example deblending. \emph{Top row, left to right:} \scarlettwo reconstructed scene without PSF, convolved with the PSF, the HSC observation, and the residuals. \emph{Rows 2--5, left to right:} the individual model for each source, its contribution to the PSF-convolved scene, the location in the observation, and its SED.}
    \label{fig:sample_deblend}
\end{figure*}

We run \scarlettwo with the neural network prior on an example scene from the HSC dataset shown in \autoref{fig:sample_deblend}, where we the HSC pipeline found four distinct peaks. The reconstructions for these sources, the entire scene with PSF convolution, and the residuals are shown in the top row of \autoref{fig:sample_deblend}. We see that there are three strongly blended sources (0,1, and 3), while source 2 is mostly isolated. Rows 2--5 of \autoref{fig:sample_deblend} show, in separate columns, the individual reconstructions of each source, the PSF-convolved source, the observation, and the spectral energy distribution (SED) of the source, respectively. Despite the strong overlap, each source morphology appears realistic, free of artifacts, and contributes to a high-fidelity reconstruction of the entire scene. For this examples, \scarlettwo required 88 iterations to converge to a relative error tolerance of $10^{-4}$ which took $\sim \ 3$ seconds.

\subsection{Ground truth testing}
\label{sec:ground_truth}

While it is nice to know that \scarlettwo yields realistic results, we want to know if it yields \emph{accurate} results.
We now test the performance of the prior-informed model and compare it with models of the same simulated scenes from \scarlet. We employ three performance metrics for flux, morphology and SED recovery: the relative error on the absolute flux
\begin{equation}
\label{eqn:flux_error}
    e_F = \frac{ \left(f - f_{\mathrm{true}}\right)}{  f_{\mathrm{true}} } , 
\end{equation}
where $f$ is the total flux of the scene, $f=\sum_b^\mathrm{bands}\sum_i^\mathrm{pixels}\mathbf{Y}_{b,i}$, and the correlation coefficient
\begin{equation}
\label{eqn:correlation}
    \xi_{\nu} = \frac{ \nu_{true} \cdot \nu }{ \sqrt{ \nu_{true} \cdot \nu_{true} } \cdot \sqrt{ \nu \cdot \nu } } ,
\end{equation}
where $\nu = \mathbf{S}\equiv\sum_b^\mathrm{bands}\mathbf{Y}_{b,i}$ denotes the morphology correlation and $\nu = \mathbf{A}\equiv \sum_i^\mathrm{pixels}\mathbf{Y}_{b,i}$ the SED correlation, respectively.

We create a mock catalog of 2,000 LSST-like blended images with the \textsc{BlendingToolKit} \citep{BTK}, which is based on the heavily tested \textsc{GalSim} framework \citep{2015galsim}. We chose a maximum magnitude of $i \leq 27$  and between 2 to 8 sources for each blend. Following \citet{scarlet_2018}, we plot our metrics as a function of the true source flux or the blendedness
\begin{equation}
\label{eqn:blendedness}
    \beta_k = 1 - \frac{\mathbf{S}_k \cdot \mathbf{S}_k}{\mathbf{S} \cdot \mathbf{S}_k},
\end{equation}
 where $k$ represents the $k$-th component of the scene, $\mathbf{S}_k$ its morphology images, and $\mathbf{S}=\sum_k\mathbf{S}_k$, all of which are computed from the truth images.
 
\begin{figure*}[t]
    \centering
    \includegraphics[width=1\textwidth]{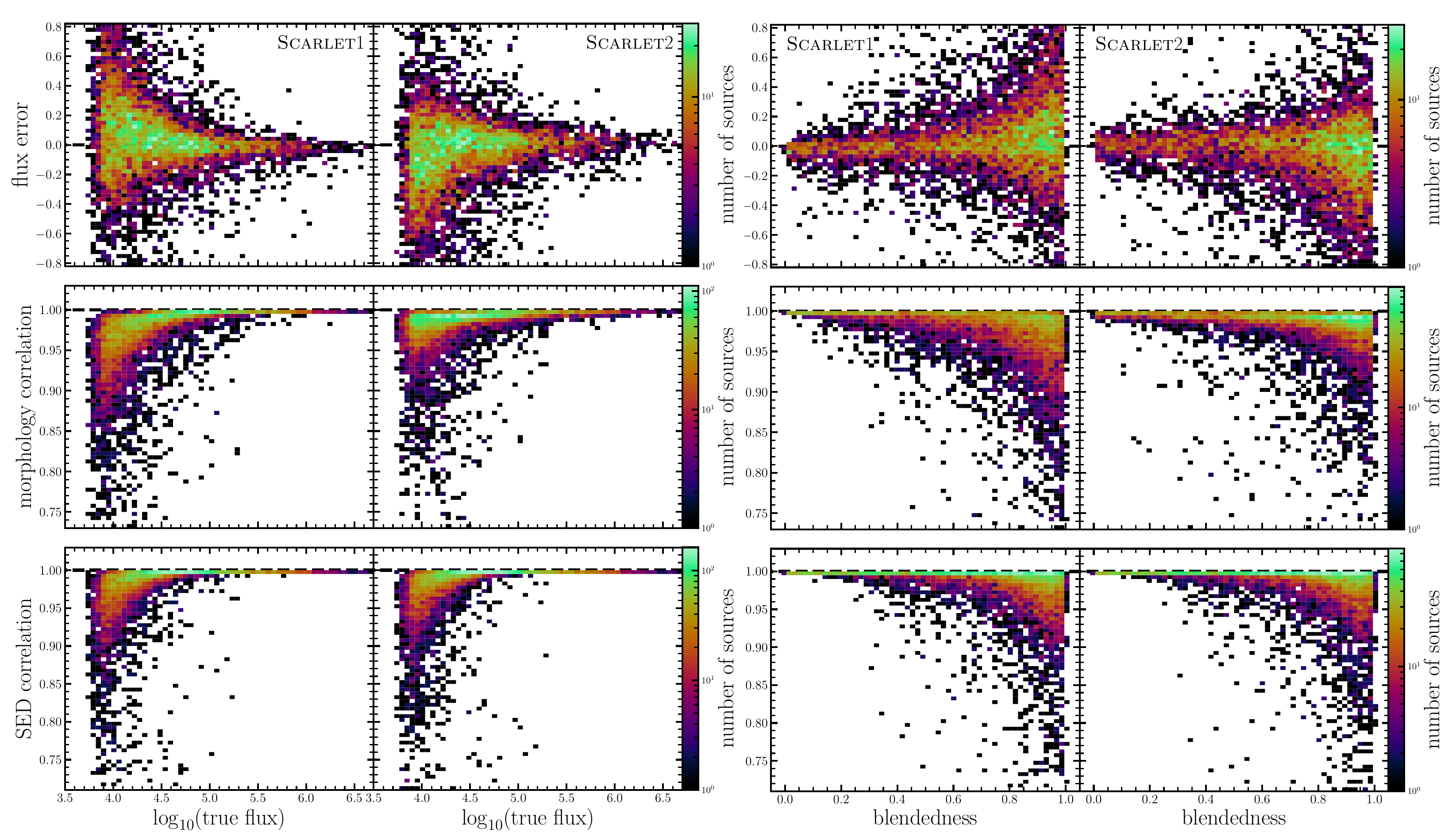}
    \caption{Comparison of deblending performance between \scarlet and \scarlettwo . From top to bottom: the relative error of total fluxes as calculated by \autoref{eqn:flux_error}, the correlation coefficient as defined in \autoref{eqn:correlation} for the morphology, and the correlation coefficient for the SED.}
    \label{fig:truth_flux}
\end{figure*}

 One would expect that fainter, and/or more blended sources will have less accurate reconstructions, and, indeed, \autoref{fig:truth_flux} show these main trends. More importantly, in both the flux (top rows) and morphology (middle rows) we see a clear improvement in accuracy when using the \scarlettwo over \scarlet. The differences are most pronounced in regions of low total flux (left side of left panels in \autoref{fig:truth_flux}), and high degrees of blending (right side of right panels in \autoref{fig:truth_flux}). The SED accuracy is very similar across both codes, which is not entirely surprising because consistent colors can be found regardless of the assumed morphology.

\subsection{Robustness to initialization}
\label{sec:init}
Due to the nature of the deblending problem, the surface of  the loss function is highly non-convex. This means that optimizers are in general sensitive to specific hyper-parameters and initializations. As explained in \autoref{sec:methods}, the proximal optimization of \scarlet is expected to be more vulnerable because paths through infeasible regions are impossible. To enable, as best as possible, convergence to a reliable model, it therefore may be necessary to have a good initialization of the source models. However, in blends sufficiently good initialization may be impossible and, even when done well, finding the optimal solution is still not guaranteed. It is therefore critical to have the sources models be robust to inaccurate initialization.

\begin{figure*}
    \centering
    \includegraphics[width=0.88\textwidth]{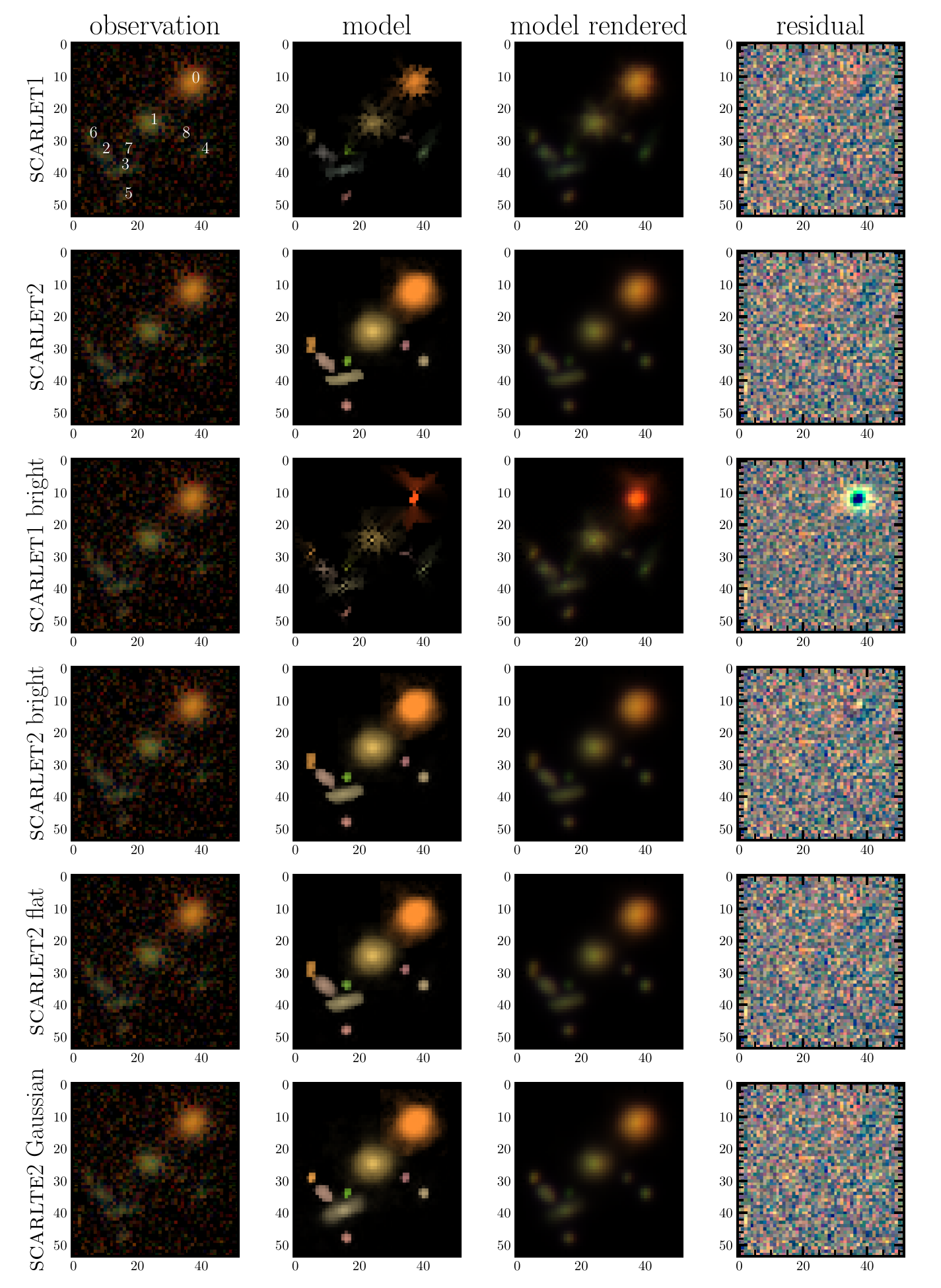}
    \caption{We show comparisons between the visual deblending performance of \scarlet and \scarlettwo on a set observation. We change the initialization of the SED from its best-fit values (in \textbf{top 2 rows}) to 20 times brighter in each band (\textbf{rows 3 and 4}). In \textbf{row 5} we show results from initializing the spectrum of \scarlettwo with a flat array of ones for each band and \textbf{row 6} we show the result from initializing each source morphology with a 2D Gaussian. }
    \label{fig:bad_init}
\end{figure*}

\autoref{fig:bad_init} shows the results of the source reconstruction of models with intentionally poor initializations for the multi-band spectrum or the morphology. Specifically, we take the observation shown in the leftmost panels, and use \scarlet to estimate the spectrum and morphology of each identified source.
We also run \scarlettwo with the same initialization as \scarlet.
The models are shown in the top two rows of \autoref{fig:bad_init}. 

We then multiplied the spectrum values by a factor of 10, resulting in an artificially bright initialization. We see in the row 3 of \autoref{fig:bad_init} that \scarlet performs poorly in this setup, with obvious errors in the models and overall a poor fit to the data. While the scenario appears contrived, thus failure mode has been observed in practice when users unintentionally provided wrong flux estimates. \scarlettwo fares much better when confronted with such a poor initialization and produces models that are virtually indistinguishable from the earlier better initialization (row 4). Even with a flat spectrum initialization, the \scarlettwo model is still practically unchanged. All trials were run for 200 iterations.
But we point out that running for even more iterations would not have improved the \scarlet models, demonstrating that the proximal optimizer is indeed trapped, not just slower than the prior-augmented version.

\begin{figure*}
    \centering
    \includegraphics[width=1\textwidth]{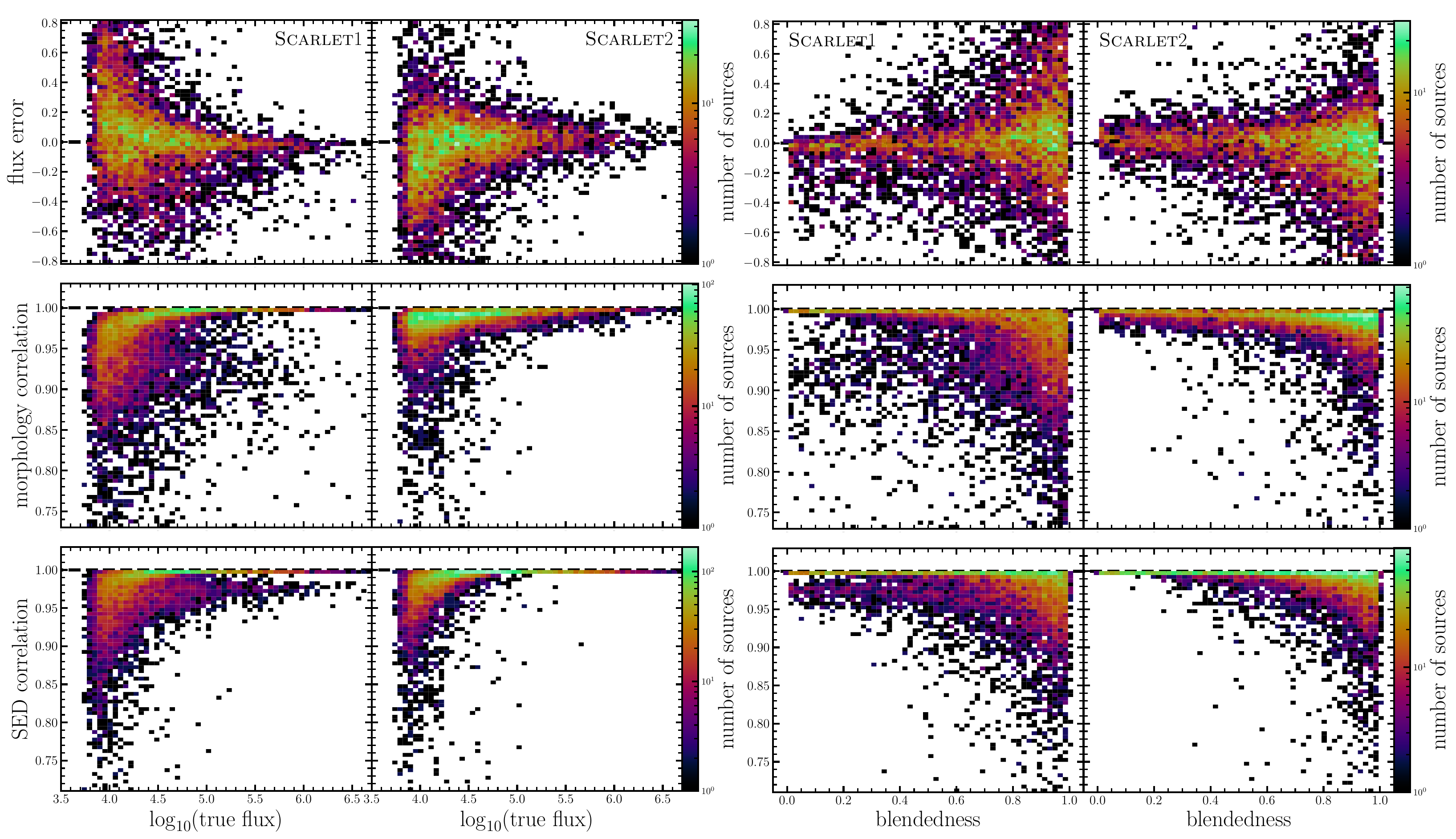}
    \caption{Same as \autoref{fig:truth_flux}, but with the flux initialized too high by a factor of 20.}
    \label{fig:poor_flux}
\end{figure*}

For a more rigorous analysis of the robustness to SED initializations,
we also run the test of \autoref{sec:ground_truth}, but this time with the initialization for the SED set to 20 times the initial value calculated by \scarlet. We again allow for both \scarlet and \scarlettwo to run for up to a maximum of 200 iterations, with a relative step size taken for both the morphology and SED parameters which results in the best case fitting for both deblending codes. \autoref{fig:poor_flux} shows the deblending accuracy metrics. We see quite notable differences in flux error, and morphology and SED correlations for the \scarlet trials, with large amounts of scatter around the optimal values for all metrics. Notably, even for bright and less blended sources, \scarlet struggles, with the SED correlation that never reach 1. This is the same problem seen in \autoref{fig:bad_init}, namely that sufficiently large changes in the overall flux lead to prominent errors in color and morphology. \scarlettwo, on the other hand, does an excellent job recovering from these poor initializations: there is little difference between the performance of the deblender with a good and an intentionally bright SED initialization.

\begin{figure*}
    \centering
    \includegraphics[width=1\textwidth]{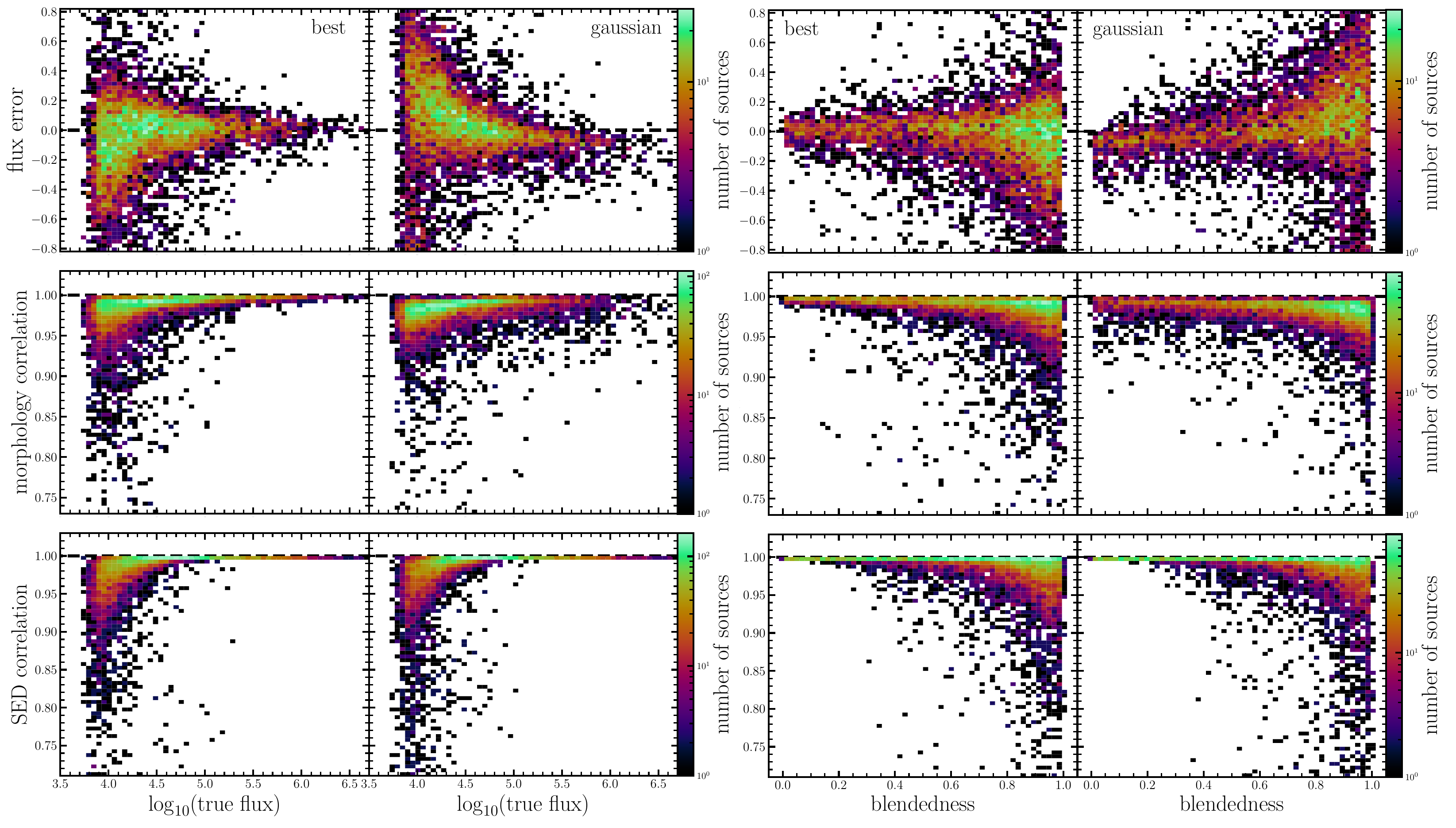}
    \caption{Same as \autoref{fig:truth_flux}, but with the left columns of each 2 panels showing \scarlettwo with the best fit initialization, and the right with a 2D Gaussian.}
    \label{fig:morph_test}
\end{figure*}

We also test the performance of \scarlettwo with random morphology initializations. \scarlettwo is ran once with the best fit initializations, and once with each source being initialized by a 2D Gaussian roughly $20\%$ of the box size drawn around the center of each source. This is the same initialization used in row 6 of \autoref{fig:bad_init}. The results are shown in \autoref{fig:morph_test} with the best-fit initializations in the left columns and the 2D Gaussians in the right. We see a mostly similar performance of both initializations, with the differences being most prominent at both the brightest and least blended, and faintest and most blended ends. The 2D Gaussian initialization performs well for faint and/or highly blended sources, which is most clear in the middle panels showing the morphology correlation. For brighter, less blended sources the best-fit initializations are marginally better for the morphology, however, this is not completely unexpected as these sources are likely to be larger, and less circular in shape, meaning the optimizer would need to do make larger changes when initialized with the Gaussian. Both trials have near identical results in the SED correlations. Interestingly, the flux of faint sources tends to be overestimates when using the Gaussian initializations, as compared to slightly underestimates with the best-fit model.

Due to the extra work required by the optimizer, we run the Gaussian initialized trials with a \textit{cosine one-cycle} scheduler \citep{Smith2017-bh}, as implemented in \texttt{optax}, which scales the step sizes taken. This allows for smaller initial steps during the initial phase when the confidence in the gradient direction is weaker, and larger steps during the middle stages of the optimization routine.

\subsection{Robustness to choice of prior}
\label{sec:priors}
Learning the prior is in itself complicated because one needs access to a sufficient number of deblended and deconvolved galaxy images that have the same properties as those found in the target observations. But the availability and fidelity of such images is often limited, e.g. to regions with space-based data, where blending and PSF convolution can be ignored. In our case, we trained the prior on earlier \scarlet models of HSC observations. While usually very accurate, we know that \scarlet does not always obtain perfect deblended models, and there is a tendency for structures such as radial streaks to appear on the outsides of these models due to the limitations of proximal optimization.
This begs the question how accurately the prior needs to be in order to be useful.

\begin{figure*}
    \centering
    \includegraphics[width=1\textwidth]{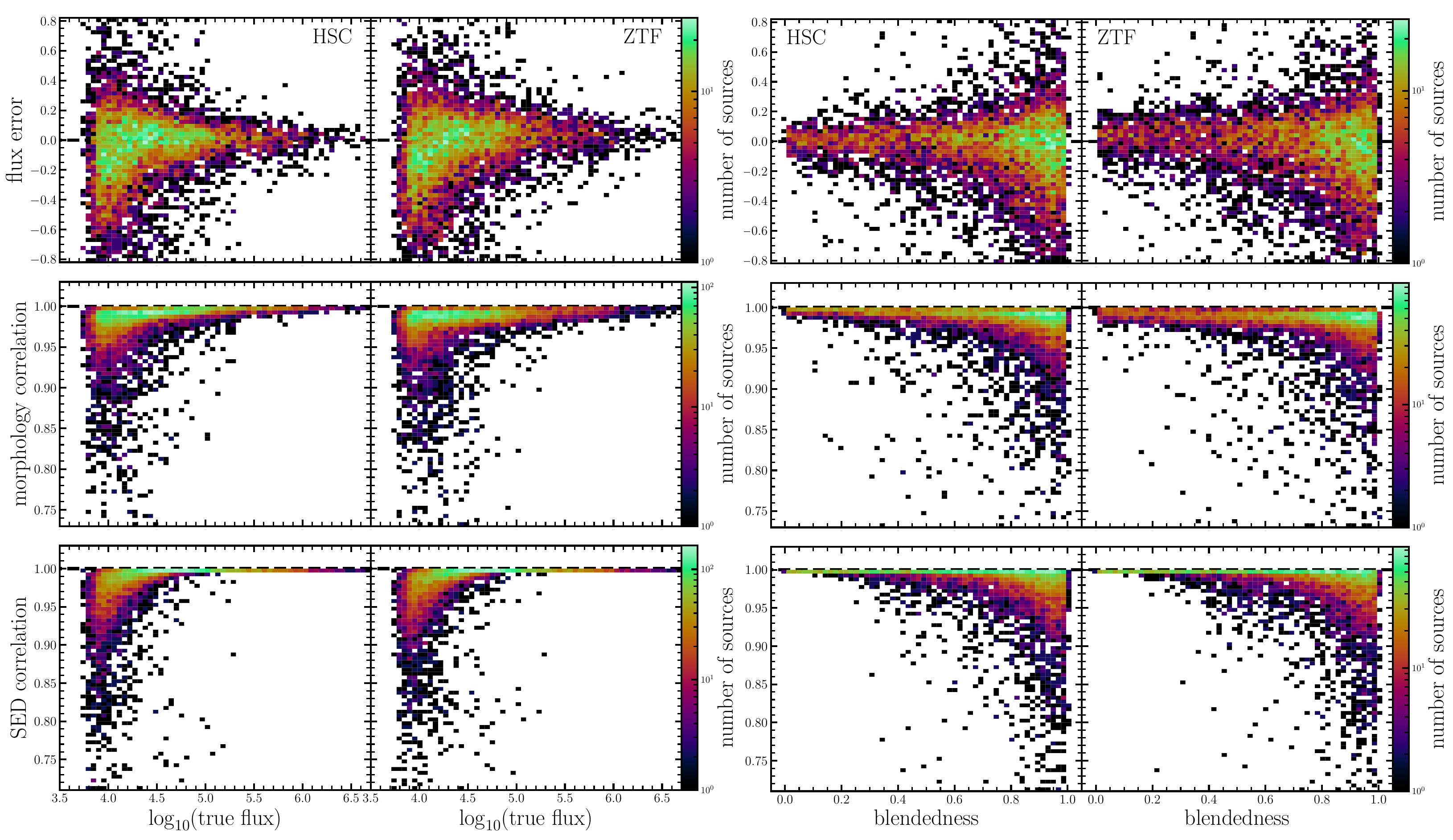}
    \caption{Same as \autoref{fig:truth_flux}, but with the left columns of each 2 panels showing \scarlettwo with the HSC prior, and the right with a ZTF prior.}
    \label{fig:prior_test}
\end{figure*}

We test the performance of \scarlettwo on HSC-like examples, but with an ill-fitting prior, specifically, the ZTF prior discussed in \autoref{sec: dataset}. It was made from simulated galaxies with properties characteristic of brighter AGN hosts. \autoref{fig:prior_test} shows the results of this test, where we observe very similar performance across the flux error and SED, with the morphology being slightly worse performing for bright and/or weakly blended sources when using the ZTF prior.
This test demonstrates that good deblending can be achieved with \scarlettwo even if the prior is only approximately correct. We interpret this result as the prior providing some stability for the optimizer to choose and maintain broadly correct update directions, with the detailed inadequacy of the prior being hidden by the noise.
Similar to the approaches of \citet{Rozet2024-gm} and \citet{Barco2024-bz}, we will in the future investigate if applying a sufficiently accurate prior will allow us to fit a wide range of data, from which we can then learn an improved prior that is less affected by inherent biases in the original prior.

\subsection{Performance summary}
\begin{figure*}
    \centering
    \includegraphics[width=1\textwidth]{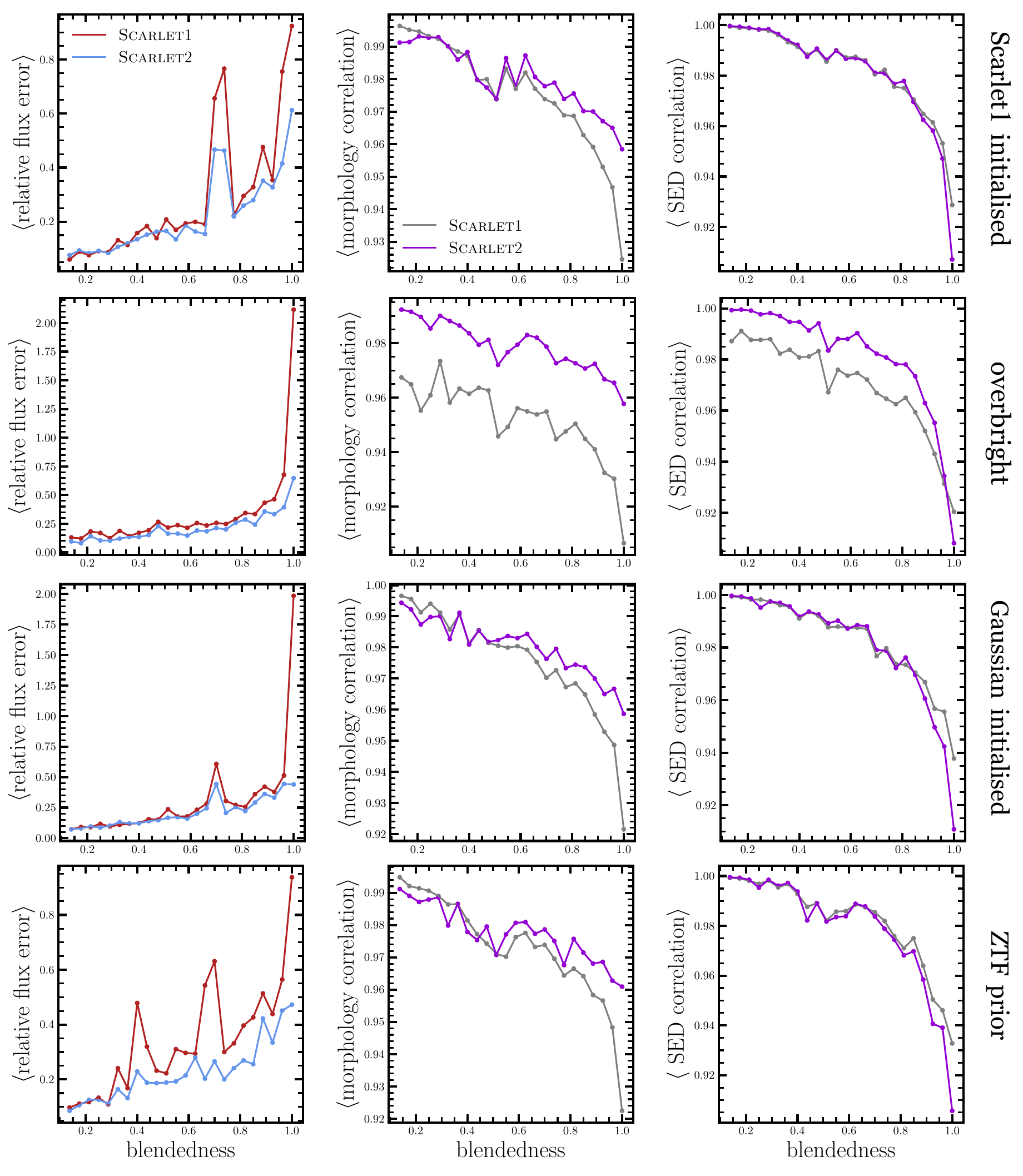}
    \caption{Summarised data from Figures \ref{fig:truth_flux}, \ref{fig:poor_flux}, \ref{fig:morph_test}, and \ref{fig:prior_test} from top to bottom rows respectively. The panels in each row show the relative flux error, morphology correlation, and SED correlation all as a function of blendedness from left to right respectively. We note the change of colour to emphasise in the left column, a lower value represents a better fit, whereas in the right two columns the better fit is shown by the higher values.}
    \label{fig:comparisons}
\end{figure*}

Overall we find that using a score based prior improves the robustness, and accuracy of the deblending algorithm across a range of test scenarios as shown in Figures \ref{fig:poor_flux}, \ref{fig:morph_test}, and \ref{fig:prior_test}. We show an averaged version of these histograms plotted as a function of blendedness (\autoref{eqn:blendedness}) for more clear by eye comparisons in \autoref{fig:comparisons}. We see that the \scarlettwo algorithm performs better throughout in terms of relative flux error (column one), with \scarlet performing notably poor on extremely blended sources. The morphology correlations as stated previously show that the prior helps significantly when poor initialisations are taken for the spectrum, with again most significantly discrepancies at high blendedness values. The SED correlations as seen in the rightmost column of \autoref{fig:comparisons} are quite similar with the lone exception being the improved performance of \scarlettwo when a poor initial spectrum (row two) is chosen.

\subsection{Quantifying hallucinations}
\label{sec:hallucinations}
\begin{figure*}
    \centering
    \includegraphics[width=0.97\textwidth]{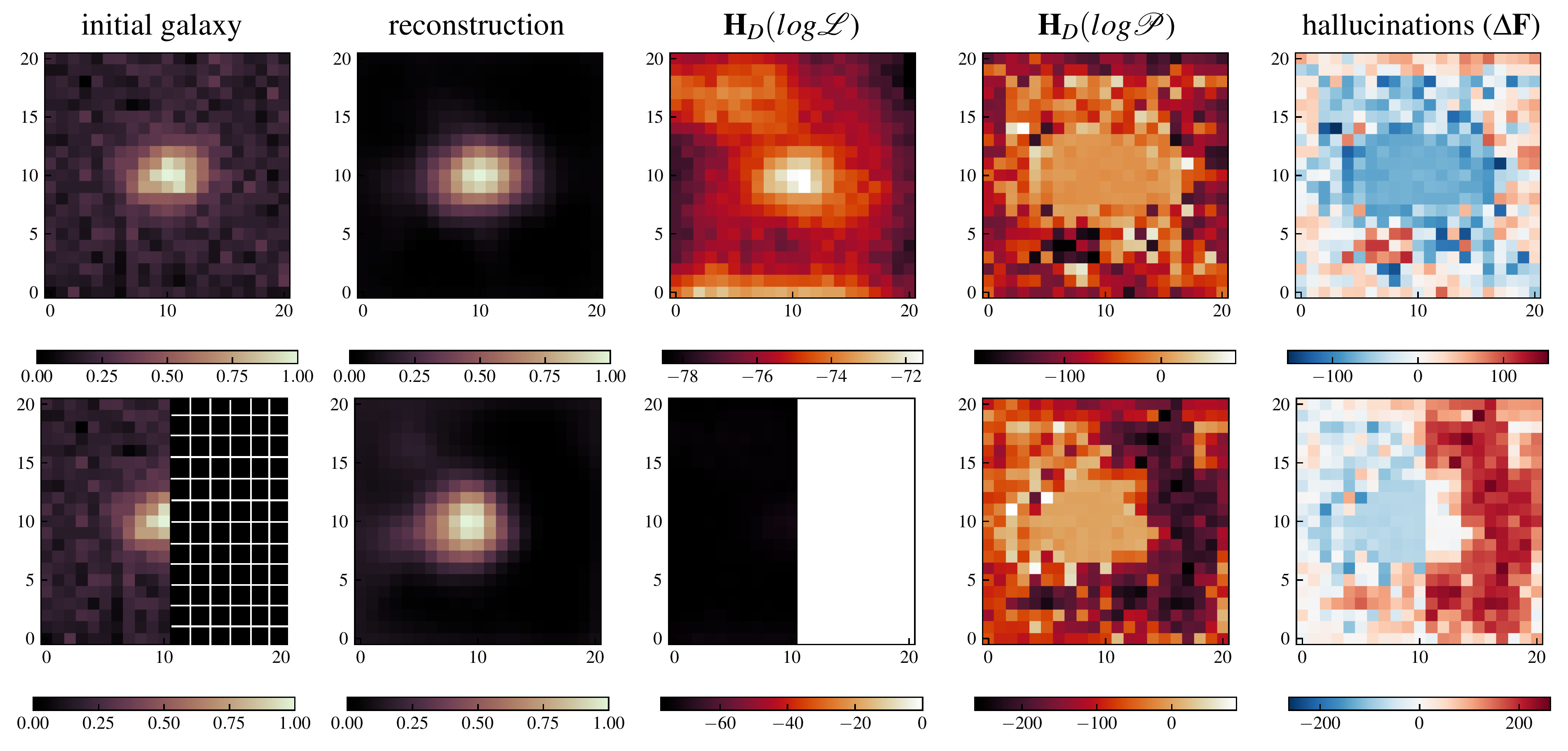}
    \caption{Hallucination score. The first two panels show the input images and the reconstructed \scarlettwo models. Panels 3 and 4 show the Hessian diagonal for $\log \mathcal{L}$ and $\log \mathcal{P}$, respectively. We note the features of $\mathbf{H}_D (\log \mathcal{L})$ come directly from the variance weighting in the initial HSC data for the galaxy. The rightmost panel show the hallucination score $\Delta \mathbf{F}$ from \autoref{eqn:hallucinate}. The red shading indicates regions dominated by the prior, whereas the blue shading shows regions dominated by the likelihood. In the second row, the input image is identical, but the right half is masked out.}
    \label{fig:hessian2}
\end{figure*}

An inevitable problem when using neural networks to solve inverse problems is the possibility of \textit{hallucinations} appearing in the reconstructed images. In this context, hallucinations are defined as features in the reconstructed image that come directly from the prior as opposed to the data. Particularly in low SNR regions or regions of significant blending the constraining power of the data is severely limited, so that the posterior will necessarily be prior-dominated. While we have demonstrated in \autoref{sec:scarlet2_metrics} that the addition of a prior improves the reconstruction of blended sources, it behooves us to provide a measure of (un)ambiguity of the results, i.e. we want to quantify what areas of the source model are constrained by the prior rather than the likelihood.    

In \citet{sampson_hallucinate_23}, we presented a practical approach. In short, we define a \textit{hallucination score} as
\begin{equation}
\label{eqn:hallucinate}
    \Delta \mathbf{F} = \mathrm{Diag}(\mathbf{F}_{\log \mathcal{P}}) - \mathrm{Diag}(\mathbf{F}_{\log \mathcal{L}}),
\end{equation}
where $\mathrm{Diag}(\mathbf{F})$ is the diagonal of the Fisher information matrix, i.e. the negative Hessian of the logarithmic probability, calculated separately for the prior and the likelihood. While the Hessian of the latter can be computed with automatic differentiation, for the former we use Hutchison's method to exploit that we already have the first derivatives in the form of the score model $s_\theta$ (see \citet{sampson_hallucinate_23} for full details).

\autoref{fig:hessian2} shows an example of this method with the diagonal Hessians for both the likelihood and the prior, as well as the hallucination score. In the bottom row, we artificially set the value of the image weights in the right half of the scene to zero, meaning we have no information on the galaxy morphology in this region. As expected, the hallucination score of the second case reveals that the right side of the source model is strongly prior-dominated. In both cases, the outskirts of the galaxy model are more due to the prior than the inner regions where the data is more significant.

To form an overall prior-vs-likelihood metric, we define the scalar confidence 
\begin{equation}
\gamma = -\Delta \mathbf{F} \cdot \mathbf{S},
\end{equation}
i.e. the hallucination score weighed with the morphology of the source, where brighter regions matter more for the fidelity of the model.
The confidence $\gamma$ can be seen as a generalization of conventional SNR measures because for isolated sources it scales with the significance of the data, and for blended sources it incorporates the uncertainty that is caused by the overlap through the contributions from the prior.



The calculation of the hallucination score and the confidence score is fully automated in \scarlettwo, and can be requested by the users after the optimization whenever such a detailed analysis is warranted.



\subsection{Runtime comparison}
The sheer size of data associated with upcoming large surveys places a premium on computational efficiency of any deblending tool intended to be used at scale.
Because \scarlet is already employed as deblender in the HSC and LSST pipelines, our aim for \scarlettwo is that of comparable runtimes. The main differences lie in 1) a high cost for each evaluation of the score model, 2) just-in-time compilation of \scarlettwo code, and 3) the ability to migrate the code to GPUs, whereas \scarlet is limited to CPUs.

On CPU nodes, we find that for a typical case of five sources per image and 50 optimizer iterations \scarlettwo takes $\sim 3.9$ times longer than \scarlet. When running for 200 iterations, the factor is reduced to $\sim 1.5$, which is explained by the initial overhead and subsequent gains from just-in-time compilation.
For a single source and 50 iterations, \scarlettwo takes almost an identical amount of time to run, which reveals the higher cost of score-model evaluations in \scarlettwo compared to those of proximal operators in \scarlet. On average, from testing performed in \autoref{sec:init} and \autoref{sec:priors} we find on average \scarlet takes $\sim 20-30$ steps to convergence when initialized with the best fit morphology and spectrum with \scarlettwo taking $\sim 50$ for a relative error tolerance of $1 \times 10^{-3}$. When initialized poorly, \scarlet will often run to the max iteration set, while \scarlettwo converged in $\sim 100$ iterations. More fine-tuning with optimizer parameters and scheduling may further reduce the iteration counts of \scarlettwo.


On 40GB Nvidia A100 GPU of the NERSC cluster Perlmutter\footnote{\url{https://docs.nersc.gov/systems/perlmutter/architecture/}}, a single source with 50 iterations runs on average are $\sim 1.5$ times faster than \scarlet on CPUs. For five sources, the GPU trials are $\sim 3.5$ and $\sim 4$ times faster than the CPU version for 50 and 100 iterations, respectively.

We can conclude that despite the additional score model evaluations for every source in every iteration, runtimes of \scarlettwo are somewhat but not drastically slower than those of \scarlet. The results also suggest that running larger scenes with more sources should make good use of the processing power and large memory of modern GPUs.

\section{Summary}
\label{sec:summary}

We presented an implementation of non-parametric galaxy morphology priors by a diffusion model and demonstrate its efficacy for source separation in multi-band survey data such the upcoming LSST. 
This prior is integrated into a new version of the source modeling framework \scarlet \citep{scarlet_2018}, dubbed \scarlettwo, which is implemented in \texttt{jax} and \texttt{equinox}. 
The addition of priors has allowed for the removal of the proximal operator-based constraints, which \scarlet relies on to ensure physically reasonable properties of deblended sources. 
We performed a series of accuracy and benchmarking tests for this new code, with direct comparisons made to \scarlet. Our main results are:

\begin{itemize}
    \item We find a significant improvement in the accuracy in total fluxes and morphology of deblended sources from a mock LSST dataset. The improvements are most notable in regions of either low total flux and/or highly blended sources.

    \item We find the inclusion of the data-driven prior in \scarlettwo to be very robust to inaccurate initialization. This is a significant improvement over \scarlet, whose proximal optimization is prone to convergence issues when poorly initialized.

    \item The inclusion of a data-driven prior allows for the probabilistic removal of artifacts or missing data from images. Compared to proximal projections, which simply disallow infeasible solutions, priors are more informative because they can differentiate between more and less likely image configurations.
    
    \item We find good performance of the deblending routine even when using slightly inadequate priors. This shows promise for the use of this method not only on large-scale surveys such as HSC and LSST, but also for any custom source separation tasks where the training data for the score model is limited in size or fidelity.

    \item We include an efficient method for determining a confidence metric of deblended sources based on the Hessians of the prior and the likelihood. This metric allows to identify regions of low confidence in deblended sources, or to automatically flag low confidence sources. 

    \item The runtime of \scarlettwo with a score-matching model is broadly comparable to \scarlet. This can be attributed to just-in-time compilation and the zero-temperature evaluations of the score model, a massive computational gain justifiable only because galaxy morphologies follow distributions that are sufficiently easy to model. While slower by a factor of 3--4 for typical scenes on CPUs, \scarlettwo code can be run on GPUs and TPUs, and initial testing suggests comparable walltimes and the possibility of running larger scenes simultaneously.

\end{itemize}

The trained score models used for this analysis are publicly available at \url{https://github.com/SampsonML/galaxygrad}.



\section*{Acknowledgments}
M.L.S would like to thank Roohi Dalal for her helpful comments and discussion about the scientific applications of this work. M.L.S acknowledges funding from Princeton University's first-year Fellowship in Natural Sciences.

We thank the anonymous reviewers for their comments which strengthened the paper.

\bibliographystyle{elsarticle-harv} 
\bibliography{Refs} 
\label{lastpage}
\end{document}